\documentclass[aps,showpacs,showkeys,preprintnumbers,amsmath,amssymb,nofootinbib]{revtex4}
\usepackage[colorlinks=true,linkcolor=blue,urlcolor=blue,filecolor=black,citecolor=red,pdfstartview=FitV,pdftitle={},pdfsubject={},pdfkeywords={},pdfpagemode=None,bookmarksopen=true]{hyperref}
\usepackage{graphicx}
\usepackage{amsmath}
\usepackage{amsfonts}
\usepackage{amssymb,ulem}
\usepackage{color}%
\usepackage{dcolumn}
\usepackage{MnSymbol,wasysym}
\usepackage{braket}

\newcommand{\be}{\begin{equation}}
\newcommand{\ee}{\end{equation}}
\newcommand{\bea}{\setlength\arraycolsep{2pt} \begin{eqnarray}}
\newcommand{\eea}{\end{eqnarray}}

\newcommand{\refb}[1]{(\ref{#1})}

\def\fft#1#2{{\frac{#1}{#2}}}

\def\0{{\sst{(0)}}}
\def\1{{\sst{(1)}}}
\def\2{{\sst{(2)}}}
\def\3{{\sst{(3)}}}
\def\4{{\sst{(4)}}}
\def\5{{\sst{(5)}}}
\def\6{{\sst{(6)}}}
\def\7{{\sst{(7)}}}
\def\8{{\sst{(8)}}}
\def\sst#1{{\scriptscriptstyle #1}}

\usepackage{eurosym}
\usepackage{calrsfs}
\usepackage[usenames,dvipsnames,svgnames]{xcolor}
\usepackage{color}%

\begin{document}
\title{ Nonlinear electrodynamics AdS black hole and related phenomena  in the extended thermodynamics}
\author{Xiao-Mei Kuang}
\email{xmeikuang@yzu.edu.cn}
\affiliation{Center for Gravitation and Cosmology, College of Physical Science and Technology, Yangzhou University, Yangzhou, 225009, China}

\author{Bo Liu}
\email{fenxiao2001@163.com}
\affiliation{School of Physics, Northwest University, Xi'an, 710069, China\\
School of Arts and Sciences, Shaanxi University of Science and Technology, Xi'an, 710021, China\\
Center for Gravitation and Cosmology, College of Physical Science and Technology, Yangzhou University, Yangzhou, 225009, China}

\author{Ali \"{O}vg\"{u}n}
\email{ali.ovgun@pucv.cl}
\homepage{http://www.aovgun.com}
\affiliation{Instituto de F\'{\i}sica, Pontificia Universidad Cat\'olica de
Valpara\'{\i}so, Casilla 4950, Valpara\'{\i}so, Chile}

\affiliation{Physics Department, Arts and Sciences Faculty, Eastern Mediterranean University, Famagusta, North Cyprus via Mersin 10, Turkey}

\affiliation{School of Natural Sciences, Institute for Advanced Study, 1 Einstein Drive Princeton, NJ 08540, USA}

\date{\today }
\begin{abstract}
\begin{center}
{\bf Abstract}
\end{center}
In this paper, we present the analytic solution for Anti-de Sitter-Nonlinear  electrodynamics (AdS-NED) black hole in four dimensions. We study tunneling of vector particles from the AdS-NED black hole to find Hawking temperature as well as its thermodynamical properties in usual and extended phase space. Then we explore the properties of  pressure-volume criticality by analyzing the phase diagram and its free energy. Finally, we investigate the Joule-Thomson coefficient, in which the inversion curve of warming and cooling are studied, and the ratio between the inverse temperature and critical temperature are computed in the black hole solution.

\end{abstract}

\maketitle

\section{Introduction}

Black holes are a fascinating part of our universe and also in general relativity. Recently, astronomers stumbled upon a fascinating finding: thousands of black holes likely exist near the center of our galaxy using  some of the data was collected nearly 20 years ago by NASA's Chandra X-Ray Observatory \cite{Chandra}. In near future Event Horizon Telescope (EHT) will resolve the event horizon of the central black hole using radio data \cite{eht}. Black holes are the solutions of Einstein's field equations, and asymptotically flat and stationary black holes are the simple form of them which are only showed the property of mass, angular momentum and charges. One of the big problem in general relativity is the singularities which occur in the beginning of the universe and also in black hole solutions or curvature singularities \cite{Gibbons2000mx,Novello2000km}. Maxwell theory also has similar singularities. To remove the singularities, one can use the modified Maxwell theory or nonlinear electrodynamics, which gives the Maxwell filed at weak field limits. Another interest in nonlinear electrodynamics is because of the possibility of providing quantum gravity corrections to Maxwell fields. For this purposes, large class of regular black holes are studied in literature where there is no singularities. The first one is studied by Bardeen in 1968 and known as Bardeen black hole \cite{bardeen}. Then Ayon-Beato et al. extended it using the nonlinear magnetic monopole field \cite{AyonBeato:1998ub} and it was further adjusted by Bronnikov in \cite{Bronnikov:2000vy,Bronnikov:2017sgg}. Furthermore, Dymnikova also studied nonsingular black holes that are connected with a de Sitter spacetime \cite{Dymnikova:2015hka}. Another example is the black hole solutions  found by using the noncommutative geometry \cite{Mehdipour:2016vxh}. There have recently been many studies on nonlinear electrodynamics in the black holes \cite{Kruglov:2016ymq,He:2017ujy,Nojiri:2017kex,Balart:2014cga,Cataldo:1999wr,Toshmatov:2017zpr,Fan:2016hvf,Ma:2015gpa}.

Since the Maldacena's idea of AdS/CFT correspondence \cite{Maldacena:1997re}, where the gravity theory in AdS space with a conformal field theory (CFT) living on the boundary, black hole's solutions in the AdS spacetime have became more important \cite{Witten:1998qj}.
 The AdS/CFT duality provides us to calculate the thermodynamics properties of the AdS black holes within a certain boundary of CFT in the limit of high temperature so that the AdS/CFT correspondence is an interesting tool to investigate strongly coupled field theories such as  superconductors \cite{Hartnoll:2008vx}.

Black holes are thermal systems , so that black holes have temperature and entropy \cite{Israel:1967wq,Unruh:1976db,Shankaranarayanan:2000gb,Angheben:2005rm,Kerner:2006vu,Chen:2008ra,Li:2016wgf,Kruglov:2014jya,Sakalli:2015jaa,Kuang:2017sqa,Gonzalez:2017zdz,Li:2015oia,Akhmedov:2006pg,Ejaz:2013fla,Sharif:2012xq,Ali:2007sh}. Another important development of the thermodynamics of  black holes is due to the AdS space time and phase transitions. First, Hawking-Page showed the phase transition between the Schwarzschild-AdS black hole and thermal AdS spacetime \cite{Hawking:1982dh}. Then Chamblin et al. studied the charged AdS black holes and its relation to the liquid-gas system \cite{Chamblin:1999tk}. Afterwards, extended phase space is used to investigate the renetrant phase transition and show the relation with the van der Waals fluids that provide interesting area of black hole chemistry by treating the cosmological constant as a variable thermodynamic pressure and its conjugate quantity with thermodynamic volume \cite{Kastor:2009wy,Cvetic:2010jb,Kubiznak:2012wp}. There are many application papers of studying phase transitions, $P-V$ criticality on AdS black holes in literatures \cite{Spallucci:2013osa,Delsate:2014zma,Cai:2013qga,Liang:2017vac,Zeng:2016fsb,Mo:2016ndm,Majhi:2016txt,Fan:2016rih,Miao:2016ipk,Zeng:2016aly,Mandal:2016anc,Chen:2016gzz,Ma:2016aat,Liu:2016uyd,Poshteh:2016rwc,Guo:2016eie,Li:2016zca,Ma:2017pap,Upadhyay:2017fiw,Fernando:2016sps,Miao:2016ulg,Cheng:2016bpx,Guo:2016iqn,Liang:2016xrz,Zhang:2016yek,Dehghani:2016wmw,Karch:2015rpa,Xu:2015rfa,Sherkatghanad:2014hda,Zhang:2014eap,za1,Hendi:2017uly,Zhao:2014raa,Zhao:2013oza,Li:2014ixn,Mo:2014qsa,Kuang:2016caz,Ovgun:2017bgx,Azreg-Ainou:2014twa,Hendi:2016usw,Zhang:2016nvj,Sheykhi:2016nly}
and therein.

 Moreover, recently Joule-Thomson (JT) expansion is cleverly applied to black holes in AdS spacetime by \"{O}kc\"{u} and Ayd{\i}ner \cite{Okcu:2016tgt,Okcu:2017qgo}. The JT expansion occurs when there is a constant enthalpy which is a mass for black holes. JT expansion is used as a  isoenthalpic tool to show the thermal expansion where there is heating and cooling regimes. Note that the pressure decreases for the expanding thermal system with a temperature T so that the inversion temperature is obtained where the JT coefficient vanishes and at the inversion temperature $T_i$, and corresponding inversion pressure $P_i$, there is a cooling and heating transition \cite{Lan:2018nnp,Mo:2018qkt,Mo:2018rgq,Cisterna:2018jqg}.

Our main motivation is to present an AdS black hole solution in NED and clarify influence of the NED on the thermodynamics, P-V criticality and JT expansion of the AdS-NED black hole.

This paper is organized as follows. In Sec.II, we construct the AdS black hole using the nonlinear electrodynamics by introducing the model of NED Lagrangian similar to that in \cite{yashar}, and solve the corresponding energy-momentum tensor within the Einstein's field equations. In Sec. III, we calculate Hawking temperature using the complex path (or Hamilton-Jacobi) approach by the tunneling spin-1 vector particles from the  AdS black hole. In Sec. IV, we investigate  the thermodynamics properties of the  black hole in an usual and extended phases. In Sec. V, we study the P-V criticality and Gibbs free energy to fix the phase diagram. In Sec. VI, we investigate the JT expansion and isenthalpy curve of the black hole . We conclude the manuscript in Sec. VII.

\section{Black Hole Solution in NED}
We consider the following action
\be S=\fft{1}{16\pi G}\int \mathrm{d}^4x\sqrt{-g}\, \Big(R-2 \Lambda-\mathcal{L}(\mathcal{F})\Big) \,,\ee
where the cosmological constant $\Lambda=-3\ell^{-2}$ with $\ell$ the AdS radius. $\mathcal{L}$ is a function of $\mathcal{F}$  and  we define $\mathcal{F}\equiv F^2$
while  $F=dA$ is a field strength of the electromagnetic field. Then the covariant equations of motion deduced from the above action are
\be
\qquad \nabla_{\mu}\Big( \mathcal{L}_{\mathcal{F}} F^{\mu\nu}\Big)=0 \,, R_{\mu\nu}-\fft 12 (R-2\Lambda) g_{\mu\nu}=T_{\mu\nu}\,,
\label{eom}
\ee
where $\mathcal{L}_{\mathcal{F}}=\fft{\partial\mathcal{L}}{\partial\mathcal{F}} $ and the energy momentum tensor is
\be T_{\mu\nu}=\fft 12 g_{\mu\nu} \mathcal{L}-2\mathcal{L}_{\mathcal{F}} F_{\mu\nu}^2  \,.\label{energymomentum}\ee

In our construction, we consider the Lagrangian density for NED

\be \mathcal{L}(\mathcal{F})=-\,{\frac {\ln  \left( 1-{\alpha}^{2}\sqrt {\mid\mathcal{F}\mid} \right) }{2{\alpha}^{4}
}}-\,{\frac {\sqrt {\mid\mathcal{F}\mid}}{2{\alpha}^{2}}}
,\label{lagrangian1}\ee
We note that in a weak field limit, i.e., $\alpha\to 0$, the vector field becomes $\mathcal{L}\sim \mathcal{F}$, which goes back to the linear Maxwell field theory.
We are interested in the static spherical symmetric AdS black holes with NED. To this end, we consider that
 the most general ansatz of the geometry and the electromagnetic field
\be ds^2=-f(r) dt^2+\fft{dr^2}{f(r)}+r^2(d\theta^2+\sin{\theta}^2 d\phi^2),\label{ansatz}\ee
and   \be A=-Q_m \cos{\theta}d\phi, \, \ee  where  $Q_m =\fft{1}{4\pi}\int_{\Sigma_2} F$ is the total magnetic charge carried by the black hole.
Here the square of the field strength  is given by  $\mathcal{F}\equiv F^2 = \frac{  2Q_m^2}{ r^4}= \frac{  q^2}{2 r^4}$ where  $q=2Q_m$ can be treated as the  magnetic monopole charge. Moreover,
to solve the equation of motion, we consider  the redshift function as
\begin{equation}\label{f-m}
f(r) = 1 - \frac{2 m(r)} { r}.
\end{equation}
Subsequently, in the $tt$ component of Einstein equation\footnote{We use the unit $c=\hbar=8\pi G=1$.}.
\be \label{ein2}
G_{tt} + \Lambda g_{tt} = \frac{1}{2} {\cal L(F)}g_{tt} ,
\ee
we have
\be \label{gtt}
G_{tt} = - 2 \left( 1 - \frac{ 2 m(r)}{r} \right) \frac{ m'(r)} {r^2}
\ee
and
\be \label{lag2}
{ \cal L(F)} = {\frac {1}{4{\alpha}^{4}} \left[ -\frac{\sqrt{2}{\alpha}^{2}\mid q\mid}{r^2}+2\,\ln  \left( 2 \right) -2\,\ln  \left(-\frac{\sqrt{2}{\alpha}^{2}\mid q\mid}{r^2}+2 \right)  \right] }.
\ee
For simplification, we rewrite $\mid q\mid $ as $q$ in the following studies.

Considering  Eq.\refb{gtt} and Eq.\refb{lag2} into Eq.\refb{ein2}, we do the integration and obtain the function $m(r)$ as
\begin{eqnarray}
m(r) = \frac{\Lambda\,{r}^{3}}{6}-\frac {\sqrt {2}qr}{3{
\alpha}^{2}}+\frac{2r^3}{3\alpha^4}\ln\frac{\sqrt{2}r^2}{\sqrt{2}r^2-q\alpha^2}
-\frac{2^{5/4}{q}^{3/2}}{3\alpha}{\rm arctanh} \left({\frac {\alpha q^{1/2}}{2^{1/4}r}
}\right)+m_0, \end{eqnarray}
where $m_0$ is an integral constant related with the mass of the black hole. Subsequently, according to Eq.\eqref{f-m}, we get  $f(r)$ as
 \begin{eqnarray}\label{eq-f}
f(r)=1-\frac{\Lambda{r}^{2}}{3}-\frac{2m_0}{r}+\frac {2\sqrt {2}q}{3{\alpha}^{2}}+\frac{4r^2}{3\alpha^4}\ln\left(1- \frac{q\alpha^2}{\sqrt{2}r^2}\right)+\frac{2^{9/4}{q}^{3/2}}{3\alpha r}{\rm arctanh} \left({\frac {\alpha q^{1/2}}{2^{1/4}r}
}\right). \end{eqnarray}
 We note that other black hole solution for the model has also found in \cite{Tahamtan:2015bha} via an alternative way.  When $\alpha\to 0$, $f(r)$ goes back to the redshift function of
RN-AdS black hole, $f(r)=1-\frac{\Lambda r^2}{3}-\frac{2m_0}{r}+\frac{q^2}{r^2}$,  as we expect. Besides, the solution has two singularities which are $r=0$ and $r=\frac{\alpha q^{1/2}}{2^{1/4}}$, and so it is valid only for $r>\frac{\alpha q^{1/2}}{2^{1/4}}$.

Defining the location of horizon, $r_h$, which satisfies  $f(r_h)=0$, the integral constant $m_0$ is solved as
\begin{equation}
m_0=\frac{r_h}{2}-\frac{ \Lambda  r_h^3}{6}+\frac{\sqrt{2} q r_h}{3 \alpha ^2}+
\frac{2r_h^3}{3 \alpha ^4}\ln\left(1- \frac{q\alpha^2}{\sqrt{2}r_h^2}\right)+\frac{2^{5/4} q^{3/2} \rm arctanh\left(\frac{\alpha  \sqrt{q}}{2^{1/4} r_h}\right)}{3 \alpha }=M,
\end{equation}
and $M$ is the mass of our black hole solution.


\section{Tunneling of vector particles and Hawking Radiation}
In this section, we investigate the tunneling of massive vector particles from the  AdS-NED black hole and obtain the corresponding Hawking temperature via Hamilton-Jacobi equation with Wentzel, Kramers, Brillouin (WKB) approximation.  For this purpose,  we use  the Proca equation that shows the behavior of the wave function of spin-1 field $\Psi_{\nu}$ as follows \cite{Kruglov:2014jya}
\begin{equation}
\frac{1}{\sqrt{-g}}\partial_{\mu}\left( \sqrt{-g}\Psi^{\nu\mu}\right) +\frac{%
m^{2}}{\hbar^{2}}\Psi^{\nu}=0,  \label{10}
\end{equation}
where $m$ is the mass of the tunneling vector particle.  In Eq. (\ref{10}), we can define the second rank tensor as
\begin{equation}
\Psi_{\mu\nu}=\partial_{\mu}\Psi_{\nu}-\partial_{\nu}\Psi_{\mu}.  \label{11}
\end{equation}
After we use the ansatz of the WKB approximation for the spin-1 fields
\begin{equation}
\Psi_{\nu}=C_{\nu}\exp\left( \frac{i}{\hbar}\left( S_{0}(t,r,\theta
,\phi)+\hbar\,S_{1}(t,r,\theta,\phi)+\hdots.\right) \right) ,  \label{12}
\end{equation}
where $C_{\nu}=(C_{1},C_{2},C_{3},C_{4})$ are constants, $S_{0}(t,r,\theta,\phi)$  stands for the kinetic term (classical action) of the vector particles with the higher order action corrections $S_{j=1,2,..}(t,r,\theta,\phi)$. Symmetry of the spacetime provides us to use the Hamilton-Jacobi method by setting
\begin{equation}
S_{0}(t,r,\theta ,\phi )=-Et+R(r,\theta )+j\phi +k,  \label{13}
\end{equation}%
 in leading order of the action. Here  $E$\ is energy of the vector particle and $j$ stands for angular momentum of the vector
particles. Moreover, $k$ is used as a complex constant.

Solving the Proca equation (\ref{10}) on the background of the black hole within (\ref{12}) and (\ref{13}), we obtain a set of equations in the lowest order in $\hbar $:
\begin{equation}
-E\left( \partial _{r}R\right) C_{1}-\frac{E(\partial _{\theta }R)C_{2}}{%
r^{2}f}-\frac{\left[ \sin ^{2}\theta \left( r^{2}f(\partial
_{r}R)^{2}+m^{2}r^{2}+(\partial _{\theta }R)^{2}\right) +j^{2}\right] C_{4}}{%
r^{2}f\sin ^{2}\theta }-\frac{EjC_{3}}{r^{2}f\sin ^{2}\theta }=0,
\end{equation}
\begin{equation}
-\frac{\left[ -\sin ^{2}\theta f(\partial _{\theta }R)^{2}-f(m^{2}r^{2}\sin
^{2}\theta +j^{2})+E^{2}r^{2}\sin ^{2}\theta \right] C_{1}}{r^{2}\sin
^{2}\theta }-\frac{f\left( \partial _{r}R\right) \left( \partial _{\theta
}R\right) C_{2}}{r^{2}}-\frac{jf\left( \partial _{r}R\right) C_{3}}{r^{2}\sin \theta }-E\left(
\partial _{r}R\right) C_{4}=0,  \label{15}
\end{equation}
\begin{equation}
-\frac{f\left( \partial _{r}R\right) (\partial _{\theta }R)C_{1}}{r^{2}}-%
\frac{\left[ -\sin ^{2}\theta f\left( \partial _{r}R\right)
^{2}-f(m^{2}r^{2}\sin ^{2}\theta +j^{2})+E^{2}r^{2}\sin ^{2}\theta \right]
C_{2}}{r^{2}\sin ^{2}\theta }-\frac{j(\partial _{\theta }R)C_{3}}{r^{4}\sin ^{2}\theta }-\frac{(\partial
_{\theta }R)EC_{4}}{r^{2}f}=0,  \label{16}
\end{equation}
\begin{equation}
-\frac{jf\left( \partial _{r}R\right) C_{1}}{r^{2}\sin ^{2}\theta }-\frac{%
j(\partial _{\theta }R)C_{2}}{r^{4}\sin ^{2}\theta }+\frac{\left[ f(\partial
_{\theta }R)^{2}-r^{2}(-f^{2}\left( \partial _{r}R\right) ^{2}-m^{2}f+E)%
\right] C_{3}}{r^{4}f\sin ^{2}\theta }-\frac{jEC_{4}}{r^{2}\sin ^{2}\theta f}=0,  \label{17}
\end{equation}

To find the solution of these equations, we write them in a matrix form and use the property of transition to the transposed vector
$\aleph\left(C_{1},C_{2},C_{3},C_{4}\right)^{T}=0$. Then the nonzero components are
\begin{align}
\Upsilon _{11}& =\Upsilon_{24}=-E\left( \partial _{r}R\right) , \\
\Upsilon _{12}& =\Upsilon _{34}=-\frac{E(\partial _{\theta }R)}{r^{2}f}, \\
\Upsilon _{13}& =\Upsilon _{44}=-\frac{Ej}{r^{2}f\sin ^{2}\theta }, \\
\Upsilon _{14}& =-\frac{\left[ \sin ^{2}\theta \left( r^{2}f\left( \partial
_{r}R\right) ^{2}+m^{2}r^{2}+(\partial _{\theta }R)^{2}\right) +j^{2}\right]
}{r^{2}f\sin ^{2}\theta }, \\
\Upsilon _{21}& =-\frac{\left[ -\sin ^{2}\theta f(\partial _{\theta
}R)^{2}-f(m^{2}r^{2}\sin ^{2}\theta +j^{2})+E^{2}r^{2}\sin ^{2}\theta \right]
}{r^{2}\sin ^{2}\theta }, \\
\Upsilon _{22}& =\Upsilon_{31}=-\frac{f\left( \partial _{r}R\right) (\partial
_{\theta }R)}{r^{2}},
\end{align}
\begin{align}
\Upsilon _{23}& =\Upsilon _{41}=-\frac{jf\left( \partial _{r}R\right) }{%
r^{2}\sin \theta }, \\
\Upsilon _{32}& =-\frac{\left[ -\sin ^{2}\theta f\left( \partial _{r}R\right)
^{2}-f(m^{2}r^{2}\sin ^{2}\theta +j^{2})+E^{2}r^{2}\sin ^{2}\theta \right] }{%
r^{2}\sin ^{2}\theta }, \\
\Upsilon_{33}& =\Upsilon _{42}=-\frac{j(\partial _{\theta }R)}{r^{4}\sin
^{2}\theta }, \\
\Upsilon_{43}& =\frac{\left[ f(\partial _{\theta }R)^{2}-r^{2}(-f^{2}\left(
\partial _{r}R\right) ^{2}-m^{2}f+E)\right] }{r^{4}f\sin ^{2}\theta }.
\end{align}
The solution of these equation satisfying  the condition
$\det \aleph =0$ is
\begin{equation}
\frac{m^{2}\left[ \sin ^{2}\theta f(\partial _{\theta
}R)^{2}+f^{2}r^{2}\left( \partial _{r}R\right) ^{2}+f\left( m^{2}\sin
^{2}\theta r^{2}+j^{2}\right) -E^{2}r^{2}\sin \theta \right] ^{3}}{%
r^{10}\sin \theta f^{3}}=0,
\end{equation}
where the radial equation $R$ is obtained non-trivially
\begin{equation}
R=\int\pm\frac{\sqrt{E^{2}-f\left( m^{2}+\frac{(\partial_{\theta}R)^{2}%
}{r^{2}}+\frac{j^{2}}{\sin^{2}\theta r^{2}}\right) }}{f}.  \label{29n}
\end{equation}

We show the outgoing and ingoing particles by the signs of $\pm$.
In the equation  \eqref{29n}, there is a pole at the event horizon of the  AdS-NED black hole so that we use the complex path integration method to solve it. The solution of the integral becomes
\begin{equation}
ImW_{\pm }(r)=\pm \left. \frac{\pi }{\partial _{r}f}E\right\vert _{r=r_{h}}.
\label{30}
\end{equation}%
Afterwards, we can easily find the tunneling probabilities of the particles
\begin{eqnarray}
P_{emission} & =e^{-\frac{2}{\hbar}ImS_{+}}=e^{\left[ -\frac{2}{\hbar }(ImW_{+}+Im B)\right] },\\
P_{absorption} & =e^{-\frac{2}{\hbar}ImS_{-}}=e^{\left[ -\frac{2}{\hbar }%
(ImW_{-}+Im B)\right] }.  \label{27}
\end{eqnarray}
Note that black hole must absorb all the particles classically, so that we take the probability of the ingoing particles as $P_{absorption}=1$. Next, the relation becomes $ImB=-ImW_{-}$ and use the $W_{+}=-W_{-}$, tunneling rate of the massive vector particles are obtained as
\begin{equation}
\Gamma =P_{emission}=\exp \left( -\frac{4}{\hbar }ImW_{+}\right)=\exp
\left( -\left. \frac{4\pi }{\hbar \left( \partial _{r}f\right) }E\right\vert
_{r=r_{h}}\right) .  \label{28}
\end{equation}%
It is noted that $\Gamma $ equals to the Boltzmann factor. Hence the temperature of the AdS-NED black hole is
\begin{equation}\label{Temp}
T=\left. \frac{\hbar \left( \partial _{r}f\right) }{4\pi }\right\vert
_{r=r_{h}}=\frac{1}{4\pi r_h}-\frac{\Lambda r_h}{4\pi}-\frac{2r_h}{3\pi\alpha^4}+\frac{q}{3\sqrt{2}\pi\alpha^2r_h}
+\frac{\sqrt{2}(\sqrt{2}r_h^2+q\alpha^2)}{3\pi\alpha^4r_h}+\frac{r_h}{\pi\alpha^4}\ln\left(1- \frac{q\alpha^2}{\sqrt{2}r_h^2}\right).
\end{equation}

\section{Properties of  thermodynamics }
After the Hawking temperature in hand, in this section, we shall study the thermodynamical properties of the black hole.  We first focus on usual phase space and then we will study the thermodynamics in the extended phase space by defining  the cosmological constant as the pressure \cite{Dolan:2010ha,Dolan:2011xt}
\begin{equation}\label{defP}
P=-\frac{\Lambda}{8\pi}.
\end{equation}

\subsection{Thermodynamics in usual phase space}
The Hawking temperature of the AdS-NED black hole is given in \eqref{Temp} and the entropy is
\begin{equation}
S =\frac{A}{4}=\pi r_h^2.
\end{equation}
The magnetical  potential of the black holes is
\begin{equation}
\Phi=\left(\frac{\partial M}{\partial Q_m}\right)_{S}=\frac{2^{5/4}\sqrt{q}}{\alpha}\rm arctanh\left(\frac{\alpha  \sqrt{q}}{2^{1/4} r}\right).
\end{equation}
Furthermore, it is straightforward to check that the temperature also satisfies
\begin{eqnarray}
T=\left(\frac{\partial M}{\partial S}\right)_{Q_m}.
\end{eqnarray}
Therefore, the first law of the black hole
\begin{equation}
dM=TdS+\Phi d Q_m.
\end{equation}
is fulfilled. The heat
capacity can be calculated by
\begin{eqnarray}
C_{Q_m}&=&T\left( \frac{\partial S}{\partial T}\right)_{Q_m}=-\frac{2 \pi  \left(6 \sqrt{2} \alpha ^2 q-3 \alpha^4 \left(\Lambda r_h^2-1\right)+12 r_h^2\ln\left(1- \frac{q\alpha^2}{\sqrt{2}r_h^2}\right)\right)}{3 \alpha^4
   \left(\Lambda -\frac{8 q^2}{2 r_h^4-\sqrt{2} \alpha^2 q r_h^2}-\frac{2 \sqrt{2} q}
   {\alpha^2r_h^2}-\frac{4\ln\left(1- \frac{q\alpha^2}{\sqrt{2}r_h^2}\right)}{\alpha^4}+\frac{1}{r_h^2}\right)},
\end{eqnarray}
which is required to be positive for the thermodynamical stability of the black hole.

\subsection{Thermodynamics in extended phase space}
In  the extended phase space, due to the definition \eqref{defP}, the thermodynamic quantity conjugate to the pressure is  the thermodynamic volume of black hole\cite{Dolan:2010ha,Dolan:2011xt}
\begin{equation}
V=\left(\frac{\partial M}{\partial P}\right)_{Q_m,S}=\frac{4\pi}{3}r_h^3.
\end{equation}
Then the extended  first law of the thermodynamic is
\begin{equation}
dM=TdS+\Phi d Q_m+ VdP+\mathcal{A} d\alpha
\end{equation}
where $T$, $\Phi$ and $V$ are given previously and
\begin{eqnarray}
\mathcal{A}&=&\left(\frac{\partial M}{\partial \alpha}\right)_{S,P,Q}=-\frac{4\sqrt{2} q r}{3 \alpha ^3}-\frac{8 r^3 \ln \left(1-\frac{\alpha ^2 q}{\sqrt{2} r^2}\right)}{3 \alpha ^5}-\frac{2^{5/4}q^{3/2} \rm arctanh \left(\frac{\alpha  \sqrt{q}}{2^{1/4} r}\right)}{3 \alpha ^2},
\end{eqnarray}
is the conjugation of $\alpha$  introduced to fulfill first law.
Furthermore, we can derive the generalized Smarr relation of the  AdS-NED black hole is
\begin{equation}
M=2TS+\Phi Q_m-2 PV+\frac{1}{2}\mathcal{A} \alpha,
\end{equation}
in terms of the dimensional analysis.

\section{The $P-V$ criticality and Gibbs free energy}
\begin{figure}
\center{
\includegraphics[scale=0.25]{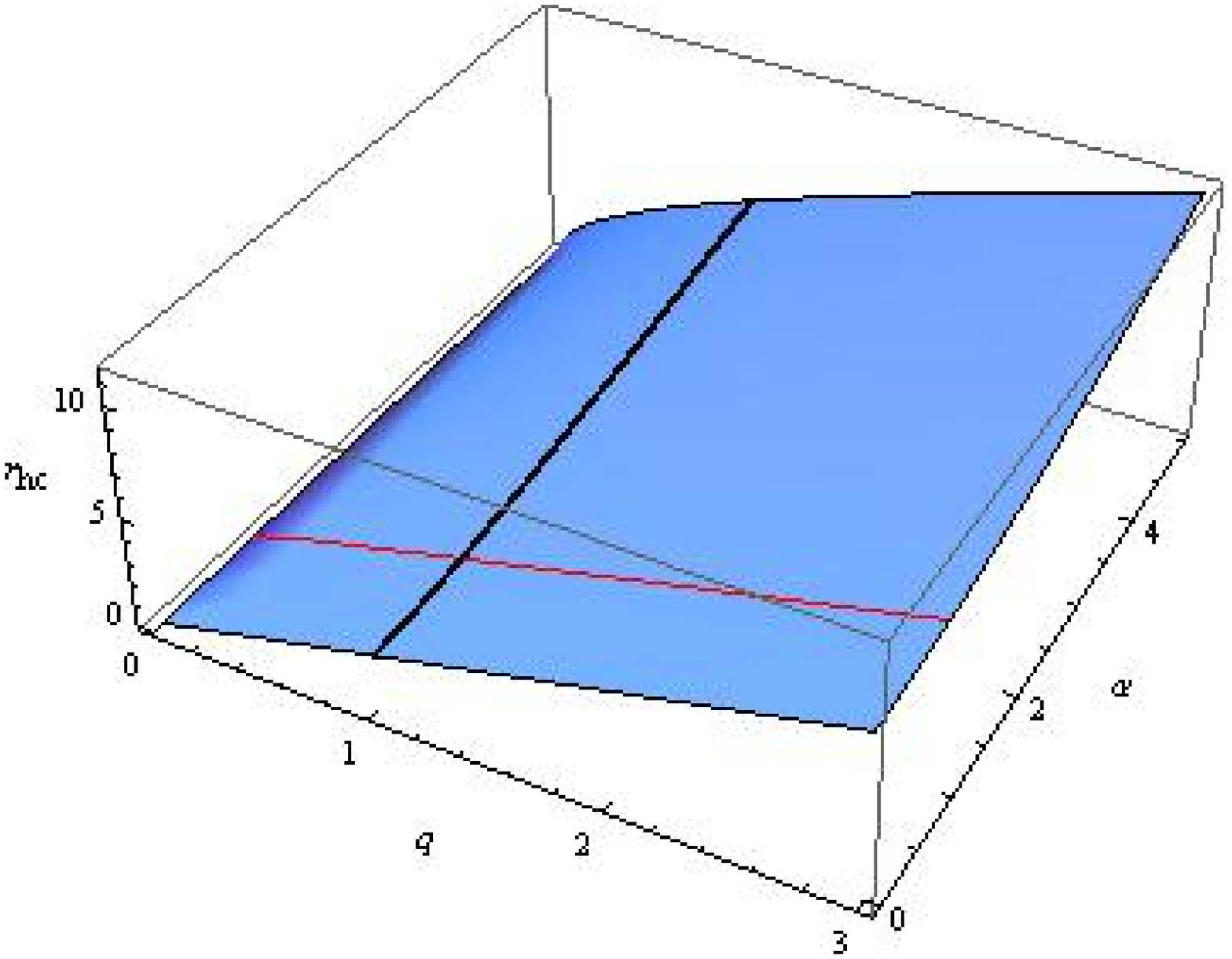}
\includegraphics[scale=0.25]{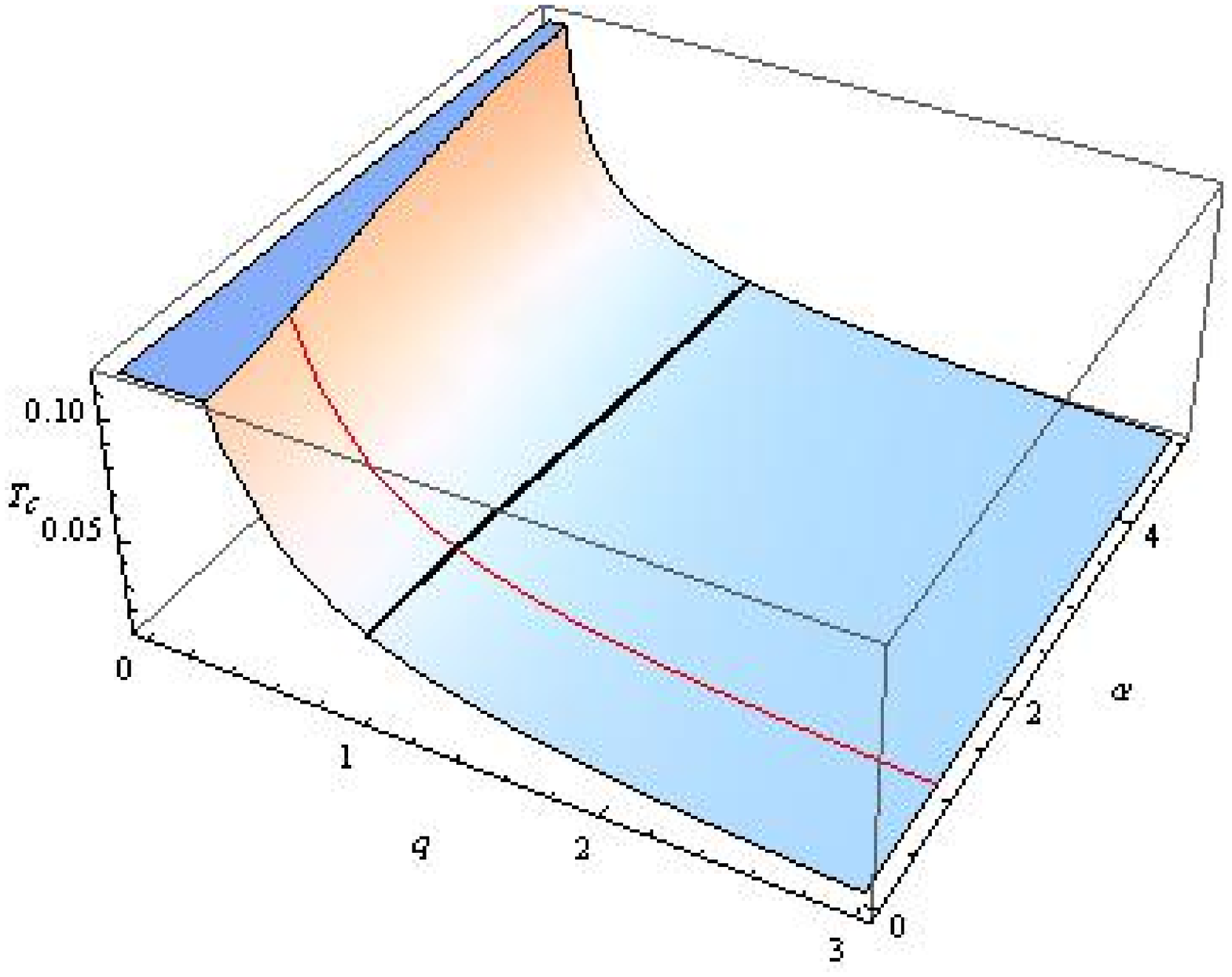}
\includegraphics[scale=0.25]{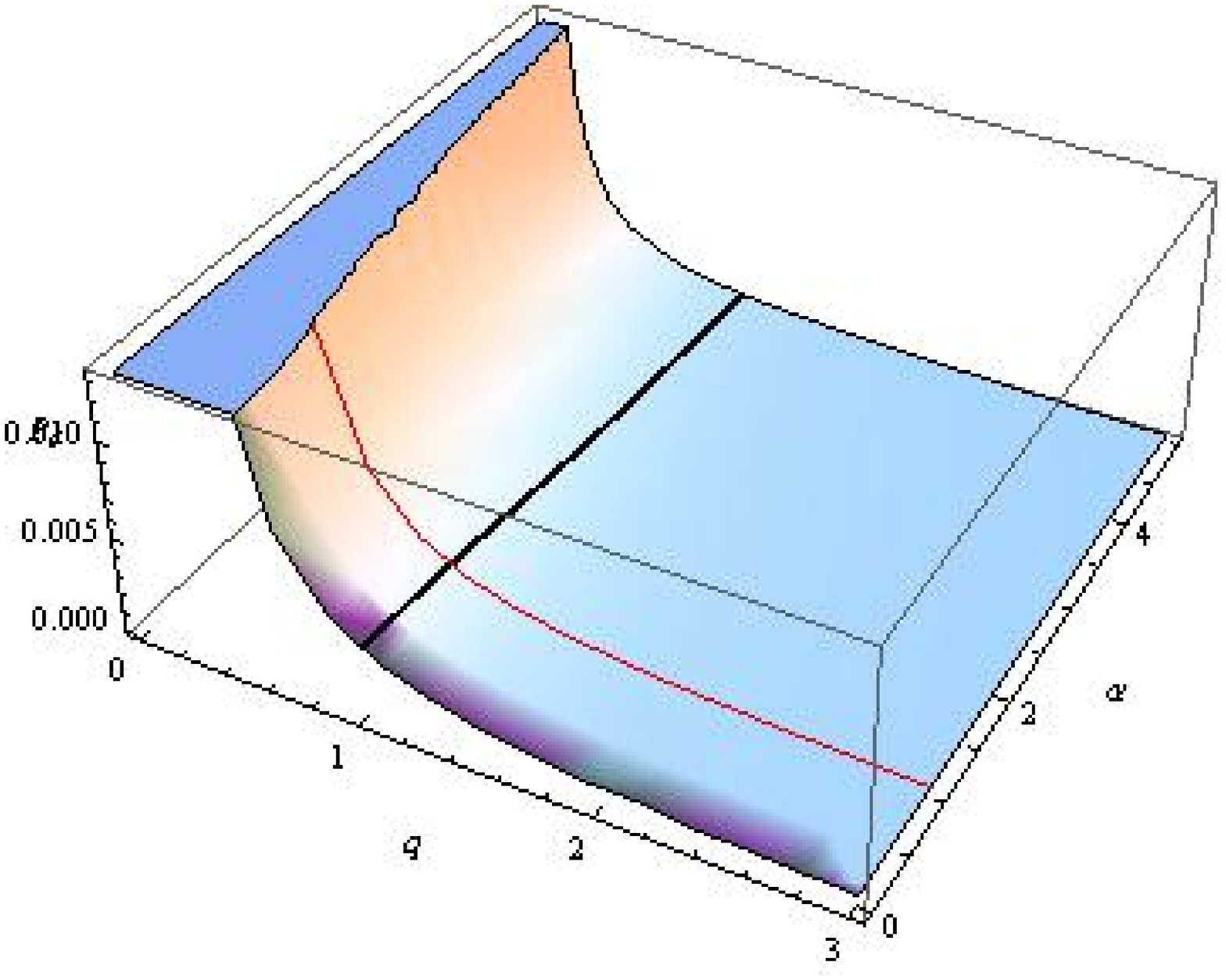}
\caption{\label{fig-criticalPoint} The dependence of critical points on the parameters.}}
\end{figure}

Then we move on to study the $P-V$ critical properties, which was proposed in \cite{Kubiznak:2012wp}, of the AdS-NED black hole  and compare with the Van der Waals fluid. To this end, we substitute the definition \eqref{defP} into the temperature \eqref{Temp}, then we reduce the state equation
\begin{eqnarray}
P=\frac{T}{2 r_h}-\frac{1}{8\pi r_h^2}+\frac{1}{3\pi\alpha^4}-\frac{q}{6\sqrt{2}\pi\alpha^2r_h^2}
-\frac{\sqrt{2}(\sqrt{2}r_h^2+q\alpha^2)}{6\pi\alpha^4r_h^2}-\frac{1}{2\pi\alpha^4}\ln\left(1- \frac{q\alpha^2}{\sqrt{2}r_h^2}\right).
\end{eqnarray}
According to the critical conditions
\begin{equation}
\frac{\partial P}{\partial r_h}=\frac{\partial^2P}{\partial r_h^2}=0,
\end{equation}
we obtain that the unique critical point is located at
\begin{eqnarray}
r_{hc}&=&\sqrt{q(X+\frac{\alpha^2}{\sqrt{2}})},\\
T_c&=&\frac{ 5 \sqrt{2} \alpha ^2 X +12\sqrt{2}\alpha ^4 q+2  \alpha ^4+36qX}{\pi  \left(3  X+\sqrt{2} \alpha ^2 \right)^2 \sqrt{4 q X+2 \sqrt{2} \alpha ^2q}},\\
P_c&=&\frac{8}{\alpha ^4}-\frac{3}{qX+\frac{\alpha ^2 q}{\sqrt{2}}}-\frac{4 \left(2 X+\sqrt{2} \alpha ^2 \right)}{\alpha ^4 X}
   +\frac{12 \left( 5 \sqrt{2} \alpha ^2 X +12\sqrt{2}\alpha ^4 q+2  \alpha ^4+36qX\right)}{\left(3 X+\sqrt{2} \alpha ^2 \right)^2 \left(2 qX+\sqrt{2} \alpha ^2 q\right)}\notag\\&&
 + \frac{12\ln(1+\frac{\alpha^2}{\sqrt{2}X})}{\alpha ^4}  +\frac{8}{X \left(\sqrt{2} \alpha ^2+2 X\right)}-\frac{4 \sqrt{2}}{\alpha ^2 \left(\sqrt{2} \alpha ^2+2 X\right)},
\end{eqnarray}
where we have defined $X=3q+\sqrt{q \left(2 \sqrt{2} \alpha ^2+9 q\right)}$. The explicit dependence of the critical points on the parameters are shown in figure \ref{fig-criticalPoint} where in the plots the red and black lines corresponds to the fixed  $\alpha=1$ and $q=1$, respectively. It is obvious that $r_{hc}$ increases as both $q$ and $\alpha$ while $T_c$  and $P_c$ has totally opposite tendency.

We note that by defining  the specific  volume as  $v=2r_{h}$\cite{Kubiznak:2012wp}, it is straightforward to obtain the relation
\begin{equation}
\frac{P_c v_{c}}{T_c}=\frac{1}{8}-\frac{\sqrt{2}X}{16\alpha^2}+\frac{3q(\frac{\sqrt{2}}{3}\alpha^2+X)\ln(1+\frac{\alpha^2}{\sqrt{2}X})}{4\alpha ^4},
\end{equation}
from which the ratio is modified by the nonlinear electrodynamic term. It can be reduced to $3/8$ as $\alpha \to 0$   for the Van der Waals fluid and predicted for any RN-AdS
black hole.

\begin{figure}
\center{
\includegraphics[scale=0.7]{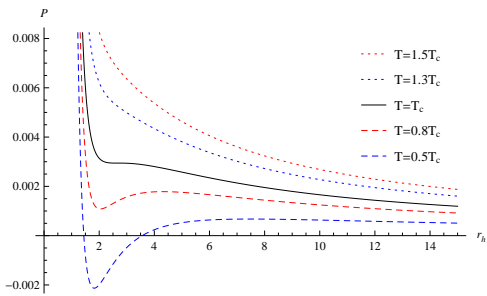}
\caption{\label{fig-Prh} $P-r_h$ diagram of NED AdS black holes.}}
\end{figure}

In the following study, we will focus on $q=\alpha=1$ without loss of generality, such that the critical points are $(T_c=0.04104, P_c=0.002943)$. We plot the behavior between the  pressure and  horizon in figure \ref{fig-Prh}. The temperature of the isothermal lines decreases from top to bottom.
The two upper dotted lines are for $T > Tc$ correspond to the ideal gas. The critical isotherm  $T = T_c$ is
denoted by the solid black line and the lower solid dashed lines correspond to
temperatures lower than $T_c$.  The $P-V$ diagram  is like the Van der Waals gas.
\begin{figure}[h]
\center{
\includegraphics[scale=0.7]{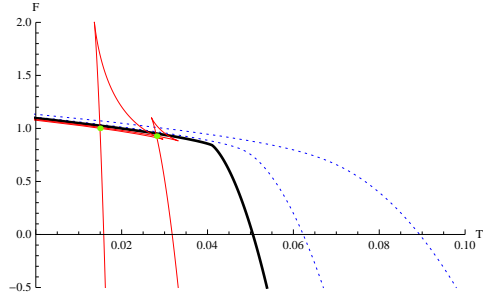}
\caption{\label{fig-TG} Gibbs free energy and the temperature. The solid black line is for the critical pressure $P_c=0.002943$ while the dotted blue lines and solid red lines are for $P>P_c$ and $P<P_c$, respectively. We set $q=\alpha=1$.}}
\end{figure}

\begin{figure}[h]
\center{
\includegraphics[scale=0.7]{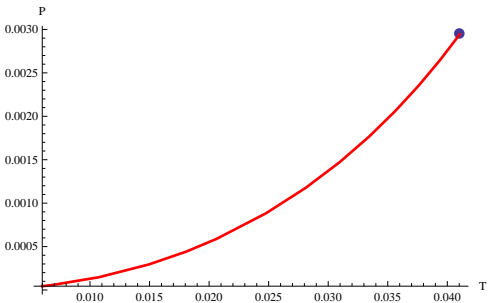}
\caption{\label{fig-CoPT} Coexistence line in $P-T$ plane. We set $q=\alpha=1$.}}
\end{figure}

Then we check the Helmholtz  free energy which is defined as
\begin{equation}
F=M-TS=\frac{2^{5/4}q^{3/2}}{3\alpha}\rm arctanh\left(\frac {\alpha q^{1/2}}{2^{1/4}r} \right)+\frac{r_h}{4}-\frac{2\pi P}{3}r_h^3
-\frac{\sqrt{2}q}{6\alpha^2}r_h-\frac{r_h^3}{3\alpha^4}\ln\left(1- \frac{q\alpha^2}{\sqrt{2}r_h^2}\right).
\end{equation}
We are interested in the fixed charged ensemble because  this ensemble exhibits many interesting features. Similar study for the fixed
potential ensemble can be straightforwardly generalized and we will not discuss
it here. The behaviour of $F$   can be seen  in figure \ref{fig-TG}. The black line is for the critical pressure $P_c=0.002943$ while the dotted blue lines and solid red lines are for $P>P_c$ and $P<P_c$, respectively. The free energy presents a swallow tail behaviour in the cases with $P<P_c$, which means a first order transition between small black hole and large black hole occurring at some $T_0$ marked with green in the figure. For each $P_0$ lower than $P_c$, there is a related $T_0$ at which two black holes have equal free energy. Then we can draw a coexistence line in the $(P, T)$-plane above which the small black hole is physically flavor while the large black hole is below the line. Our coexistence line with $\alpha=q=1$ is shown in figure \ref{fig-CoPT} where the dot denotes $(T_c,P_c)$. We note that  for the first order phase transition, the physical $P-r_h$ diagram should be
modified by replacing the oscillating part by an isobar with the transition $T=T_0$.  So the horizon the the small/large black hole with same free energy can be determined by using the Maxwell¡¯s equal area law in the $P-r_h$ diagram\cite{Spallucci:2013osa}.

\section{Joule-Thomson expansion of AdS black hole with NED}
In this section,  we investigate the Joule-Thomson expansion of the AdS-NED black hole, where occurs when a flowing gas passes through a pressure regulator, which acts as a throttling porous plug. To balance out any Joule-Thomson related temperature changes, a heating or cooling process on the AdS-NED black hole can be thought.  Joule-Thomson expansion needs a  constant enthalpy or black hole mass  in extended phase space, in other words, it is called an isenthalpic. The Joule-Thomson coefficient is defined as \cite{Okcu:2016tgt}

\begin{equation}
\label{muJ}
\mu _J =\left( \frac{\partial T}{\partial P}\right) _M=\frac{1}{C_P}\left[ T\left( \frac{\partial V}{\partial T}\right) _P-V\right].
\end{equation}

\begin{figure}
\center{
\includegraphics[scale=0.7]{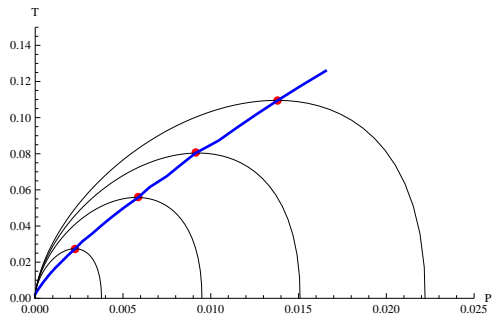}
\caption{\label{fig-PT}The isenthalpy process  in P-T graph  with fixed $M=15,18,20,22$ from bottom to top(black lines). We set $q=10$ and $\alpha=1$. The red points in each process denote the transition points of the cooling and heating regions for the related $M$, satisfying $\mu_J=0$. The blue line is $P_i-T_i$ border.}}
\end{figure}

In order to study the Joule- Thomson expansion, we rewrite the pressure $P$ as a function of the mass of black hole $M$ and the event horizon $r_h$ as
\begin{equation}
P(M,r_h)=\frac{6 \alpha ^4 M-2 \sqrt{2} \alpha ^2 q r_h-3 \alpha ^4
   r_h-4 \sqrt[4]{2} \alpha ^3 q^{3/2} \rm arctanh\left(\frac{\alpha  \sqrt{q}}{\sqrt[4]{2} r_h}\right)-4r_h^3\ln\left(1- \frac{q\alpha^2}{\sqrt{2}r_h^2}\right)}{8 \pi  \alpha ^4 r_h^3},
\end{equation}
and  the temperature is
\begin{equation}
T(M,r_h)=\frac{\alpha  \left(3 M-r_h\right)-2 \sqrt[4]{2} q^{3/2} \rm arctanh\left(\frac{\alpha  \sqrt{q}}{\sqrt[4]{2} r_h}\right)}{2 \pi  \alpha  r_h^2}.
\end{equation}
From the above equation, we can cancel the event horizon and write $T$ as a function of $(M,P)$ which is analytical difficulty. However, we numerically show the isenthalpy process with fixed $M=15,18,20,22$ from bottom to top in figure \ref{fig-PT} where we set $q=10$ and $\alpha=1$. The red points in each process denote kind of inversion point  satisfying $\mu_J=0$ we will interpret soon.
\begin{figure}
\center{
\includegraphics[scale=0.7]{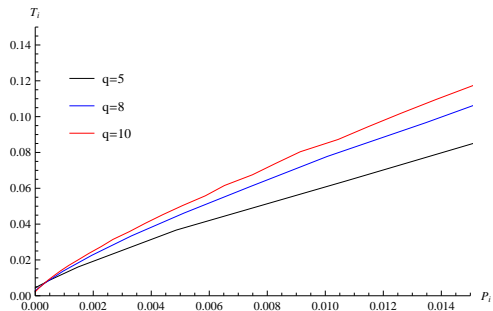}
\caption{\label{fig-Pi-Ti-q}Inversion curves $P_i-T_i$ for heating process and cooling for different charges. We set $\alpha=1$.}}
\end{figure}
\begin{figure}
\center{
\includegraphics[scale=0.7]{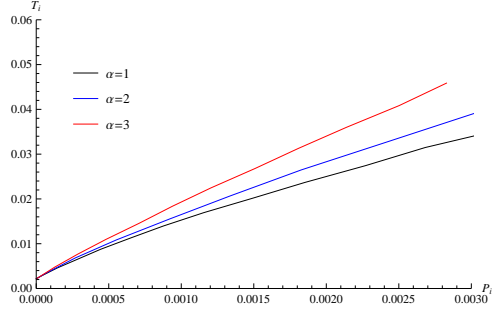}
\caption{\label{fig-Pi-Ti-a}Inversion curves $P_i-T_i$ for heating process and cooling for different $\alpha$. We set $q=10$.}}
\end{figure}

Then, we will study the properties of the inverse points with the use of  the Joule-Thomson coefficient
\begin{eqnarray}
\mu _J &=&\left( \frac{\partial T}{\partial P}\right) _M=\frac{\left(\frac{\partial T(r_h,M)}{\partial r_h}\right)_M}{\left(\frac{\partial P(r_h,M)}{\partial r_h}\right)_M}\notag\\
&=&\frac{2 r_h \left(2 \alpha  r_h \left(r_h \left(r_h-6 M\right)+2 q^2\right)+\sqrt{2} \alpha ^3 q \left(6 M-r_h\right)-4 \sqrt[4]{2} q^{3/2} \left(\sqrt{2} \alpha ^2 q-2 r_h^2\right) \tanh
   ^{-1}\left(\frac{\alpha  \sqrt{q}}{\sqrt[4]{2} r_h}\right)\right)}{3 \left(2 r_h^2-\sqrt{2} \alpha ^2 q\right) \left(\alpha  \left(r_h-3 M\right)+2 \sqrt[4]{2} q^{3/2} \tanh
   ^{-1}\left(\frac{\alpha  \sqrt{q}}{\sqrt[4]{2} r_h}\right)\right)}.
\end{eqnarray}
For $\mu_J=0$, we obtain the inversion  temperature from Eq.\eqref {muJ} as
\begin{eqnarray}
T_i&=&V\left( \frac{\partial T}{\partial V}\right) _P=\frac{4\sqrt{2}\alpha ^2 q r_h^3 \left(1-2\ln\left(1- \frac{q\alpha^2}{\sqrt{2}r_h^2}\right)\right)
+8 r_h^5\ln\left(1- \frac{q\alpha^2}{\sqrt{2}r_h^2}\right)}{12 \pi  \alpha ^4
   \left(\sqrt{2} r_h^2-\alpha ^2 q\right){}^2}\\
&&-\frac{2\sqrt{2}q\alpha^2(q^2-r_h^2)+\alpha^4q^2+8\pi Pr_h^2(2\sqrt{2}qr_h^2\alpha^2-2r_h^4-q^2\alpha^4)+ 2 r_h^4-16 \pi  P r_h^6-8 q^2 r_h^2\ln\left(1- \frac{q\alpha^2}{\sqrt{2}r_h^2}\right)}{12 \pi  r_h
   \left(\sqrt{2} r_h^2-\alpha ^2 q\right){}^2}, \notag
\label{eqJT}
\end{eqnarray}
of which the corresponding pressure is $P_i$. The relation of  $P_i-T_i$ for $q=10$ and $\alpha=1$ is shown in the blue line in figure \ref{fig-PT}, which is kind of border. In the left side of the border, the  Joule-Thomson coefficient satisfies $\mu_J>0$ which denotes a heating precess while in the right side, we have $\mu_J<0$ denoting a cooling process. From the figure, we can also see that both $T_i$ and $P_i$ increase as $M$ and only one branch exists like the observation in the previous literatures\cite{Okcu:2016tgt,Okcu:2017qgo,Lan:2018nnp,Mo:2018qkt,Mo:2018rgq}, which is different form the van der Waals fluids.
\begin{widetext}
\begin{table}[h]
\begin{center}
\begin{tabular}{|c|c|c|c|c|c|c|}
 \hline
 ~$q$ &$ P_i$&$ r_{h_i}$ & $M_i $&$ T_i^{min}$ &$T_c$&$T_i^{min}/T_c(error)$~  \\ \hline
 5& 0 &6.3762 &5.1880 & 0.004330& 0.008564& $1/2(1.13\%)$   \\  \hline
 8 & 0 &10.0520 &8.2501 & 0.002707& 0.005375& $1/2(0.72\%)$  \\  \hline
10 & 0&12.5019 &10.2915 &0.002166& 0.004306 & $1/2(0.59\%)$   \\ \hline
 \end{tabular}
 \caption{\label{table1}The related quantities for minimal inversion temperature with the change of charge. $\alpha$ is set to be $1$.}
 \end{center}
\end{table}
\end{widetext}
%
\begin{widetext}
\begin{table}[h]
\begin{center}
\begin{tabular}{|c|c|c|c|c|c|c|}
 \hline
 ~$\alpha$ &$ P_i$&$ r_{h_i}$ & $M_i $&$ T_i^{min}$ &$T_c$&$T_i^{min}/T_c(error)$~  \\ \hline
1 & 0&12.5019 &10.2915 &0.002166& 0.004306 & $1/2(0.59\%)$    \\ \hline
2 & 0 &13.2422 &10.5431 & 0.002163& 0.004234& $1/2(2.15\%)$   \\  \hline
3 & 0&14.1070 &10.9504 &0.002152& 0.004125 & $1/2(4.35\%)$    \\ \hline
 \end{tabular}
 \caption{\label{table2}The related quantities for minimal inversion temperature with the change of charge. $q$ is set to be $10$.}
 \end{center}
\end{table}
\end{widetext}

Furthermore, we show the inversion curve $P_i-T_i$ with different parameters in the NED black hole in figure \ref{fig-Pi-Ti-q} and \ref{fig-Pi-Ti-a}. In figure \ref{fig-Pi-Ti-q}, as the charge increases, the
inversion point is bigger and the curve is higher, which is consistent with that in RN case. Similar effect of $\alpha$ also can be seen in figure \ref{fig-Pi-Ti-a} which means the NED enhances the effect of the charge. In all cases, $T_i$ increases as $P_i$, so that the minimum inversion temperature $T_i^{min}$ occurs at $P_i=0$.  We list the related quantities in table \ref{table1} and table \ref{table2} with samples of $q$ and $\alpha$, respectively. Our numerical calculations show  $T_i^{min}/T_c\sim 1/2$ with ignorable errors.  This result is somehow consistent with the exact $1/2$ found by \"{O}kc\"{u} and Ayd{\i}ner \cite{Okcu:2016tgt}.

\section{Conclusion and discussion}

In this paper, we have presented an AdS black hole solution using the nonlinear electrodynamics  with new NED Lagrangian, which is converted to Maxwell's electrodynamics in weak field limit for $\alpha\to 0$. We solved the corresponding energy-momentum tensor within the Einstein's field equations to find the AdS black hole solution in four dimensions.

We have studied the thermal stability of the  AdS-NED black hole. First we have used the complex path (or Hamilton-Jacobi) approach to find the hawking radiation by the tunneling massive vector particles from the black hole. Next, we have studied the thermodynamic properties  on usual phase space, where we show that the first law of thermodynamics is satisfied and the heat capacity is obtained as positive which is mandatory for stability of the black hole. Afterwards we have presented the thermodynamical properties of the AdS-NED black hole in the extended phase space by defining  the cosmological constant as the pressure with calculating the extended first law of the thermodynamics and the generalized Smarr relation. It would be interesting to introduce the present NED action into the modified AdS gravity and study the black hole solutions as well  as its related thermodynamics.  Kinds of NED black hole solutions in higher order curvature AdS gravity and the  thermodynamics have been analyzed in \cite{Hendi:2015psa,Miskovic:2010ey} and therein.

We have also investigated the $P-V$ criticality. The $P-V$ diagram is similar to the Van der Waals fluid with fixed $q$ and $\alpha$. The critical point was analytically obtained which show that
the critical horizon increases as both $q$ and $\alpha$ while the critical temperature  and pressure  have totally opposite tendency. We also figure out the coexistence line in $(P, T)$-plane  with $\alpha=q=1$ by calculating  the Helmholtz  free energy  with fixed charged ensemble. In the left region of the coexistence line  the small black hole is physically flavor while the large black hole is stable below the line.

Lastly, JT expansion of the AdS-NED black hole have been studied.   The curve of $P_i-T_i$ with vanishing  Joule-Thomson  coefficient  was carefully analyzed.  In the left side of the curve, we have $\mu_J>0$ which denotes a heating precess while in the right side, we have $\mu_J<0$ denoting a cooling process.  Furthermore, the effect of parameters in  NED black hole on the  inversion curve was studied. And we found the NED promote the effect of the charge, which enhances the inversion points. Similar to the previous studies\cite{Okcu:2016tgt,Okcu:2017qgo,Lan:2018nnp,Mo:2018qkt,Mo:2018rgq}, the curve only presents one branch which is different form the van der Waals fluids. Finally, we studied  the ratio  $T_i^{min}/T_c$ which in around $1/2$ with tiny numerical errors. Our result is somehow consistent with the exact $1/2$ found by \"{O}kc\"{u} and Ayd{\i}ner \cite{Okcu:2016tgt}. In our model, the  study of JT expansion mainly done by numeric and the exact analytic is called for.

\acknowledgments
X.M.K. is supported by the Natural Science Foundation of China under Grant No.11705161 and Natural Science Foundation of Jiangsu Province under Grant No.BK20170481. B.L is partly supported by the Natural Science Foundation of China under
Grant nos.11675139 and 11875220. A. \"{O}. is supported by Comisi\'on Nacional
de Ciencias y Tecnolog\'ia of Chile through FONDECYT Grant N$^{\textup{o}}$ 3170035.  A. \"{O}. is  grateful to Institute for Advanced Study, Princeton for hospitality.

\end{document}